\documentclass[a4paper,fleqn,usenatbib]{mnras}

\usepackage{newtxtext,newtxmath}
  
\usepackage[T1]{fontenc}
\usepackage{ae,aecompl}

\usepackage{graphicx}	% Including figure files
\usepackage{amsmath}	% Advanced maths commands
\usepackage{amssymb}	% Extra maths symbols
\usepackage{subfigure}
\usepackage{longtable}

\title[RRLs in Circinus galaxy]{Non-detection of Broad Hydrogen Radio Recombination Lines in Circinus Galaxy}
\author[Junzhi Wang et al.]{Junzhi Wang$^{1,2}$\thanks{E-mail: junzhiwang@gxu.edu.cn}, Yong Shi$^{3}$,   Zhi-Yu Zhang$^{3}$, Shu Liu$^{4}$, Yu Gao$^{5}$,Jiangshui Zhang$^6$,
\newauthor Fengyao Zhu$^7$, and Min Fang$^8$ 
\\
$^{1}$School of Physical Science and Technology, Guangxi University, Nanning 530004, China\\
$^{2}$Shanghai Astronomical Observatory, Chinese Academy of Sciences,80 Nandan Road, Shanghai 200030, China\\
$^{3}$School  of Astronomy and Space Science, Nanjing University, Nanjing,  210093, China\\
$^{4}$CAS Key Laboratory of FAST, National Astronomical Observatories, Chinese Academy of Sciences, Beijing 100012, China\\
$^{5}$Department of Astronomy, Xiamen University, Xiamen, Fujian 361005, China\\
$^{6}$Center For Astrophysics, GuangZhou University, GuangZhou  510006, China\\
$^{7}$Research Center for intelligent Computing Platforms, Zhejiang Laboratory, Hangzhou, 311100, PR China\\
$^{8}$Purple Mountain Observatory, Chinese Academy of Sciences, 10 Yuanhua Road, Nanjing 210023, China\\
}

\date{Accepted xxx. Received xxx; in original form 2022 July 29}

\pubyear{2022}

\begin{document}
\label{firstpage}
\pagerange{\pageref{firstpage}--\pageref{lastpage}}
\maketitle

\begin{abstract}
The line widths of broad line regions (BLRs)  of  AGNs are key parameters for understanding   the central super massive black holes (SMBHs). However, due to  obscuration from dusty torus, optical recombination  lines from BLRs  in type II AGNs can not be directly detected.  Radio recombination lines (RRLs), with low extinction, can be ideal tracers to probe emission from BLRs  in type II AGNs.  We performed  RRL observations for H35$\alpha$ and  H36$\alpha$ toward  the center of Circinus galaxy with  ALMA.  Narrow components of  H35$\alpha$ and  H36$\alpha$, which are thought to be mainly from star forming regions around the nuclear region,  are detected.  However, only upper limits are obtained for broad H35$\alpha$ and  H36$\alpha$.  Since Circinus galaxy  is one of the nearest AGN,  non-detection of broad RRLs in Circinus galaxy at this band tells us that  it is hopeless to  detect broad RRL emission in local AGNs with current facilities.   Submillimetre RRLs, with flux densities that are dozens of times higher than those at the millimetre level, could be the tools to directly detect BLRs in type II AGNs with ALMA, once its backend frequency coverage has been upgraded to several times better than its current capabilities. 

%We present the first detection of a broad RRL emission from type II AGN, with the observations of H35$\alpha$ and  H36$\alpha$ toward  Circinus galaxy using ALMA.   The line width (FWHM) of H35$\alpha$ and  H36$\alpha$ broad emission is 3532$\pm$187 km s$^{-1}$, which is similar to that of H$\alpha$ with  polarized emission as $\sim$3300 km s$^{-1}$ in the literature. The upper limit of electron density in BLRs in  Circinus galaxy can be estimated to be $\sim10^{10}$cm$^{-3}$, as given by the pressure broadening of RRLs from BLRs.    Narrow H35$\alpha$ and  H36$\alpha$ emissions, which are thought to be mainly from star forming regions around the nucleus, are detected as well.  Stimulated emissions  should be important for the broad and narrow  H35$\alpha$ and  H36$\alpha$  lines in Circinus galaxy,  as suggested by the  obtained  line to continuum ratio.  

\end{abstract}

% Select between one and six entries from the list of approved keywords.
% Don't make up new ones.
\begin{keywords}
galaxies: active; galaxies: Seyfert
\end{keywords}

\section{Introduction}

Supermassive black holes (SMBHs)  reside in the centres of
most  if  not all  massive  galaxies, and  play  a key  role in galaxy
formation and  evolution \citep{2002ApJ...574..740T,2006ApJ...643..641H}.
The SMBH mass measurement is critical for understanding the growth of SMBHs and  its interplay with galaxy evolution
across  the  cosmic  time.   Active  galactic  nuclei  (AGNs)  are the
manifestation of active accretion  onto SMBHs.  
A method based  on the broad emission lines of AGN offers  the most efficient way
to  measure the  the black-hole masses  in  a large  number of  sample from  the local
universe out to the most distant  quasars at $z$=7. Such a method employs the
virial theorem of broad emission  line (BEL) regions in AGN where the velocity
is derived based on the line width of BELs and the location of BELs is
estimated  through  the  continuum  luminosity  \citep{2005ApJ...629...61K,2014SSRv..183..253P}.
 However,  this method  is  applicable only to type I  AGN where
BELs are not obscured by dusty  torus. Given the fact that type II AGN
may dominate the  AGN population, the above problem  largely limits us
from a complete understandings of many fundamental problems related to
SMBH masses, e.g., what is the  cosmic history of SMBH growth. 

 For the type II AGNs without heavy obscuration, the BELs can
be detected with near-IR observations. As a result, broad B$\gamma$ line
emission has been detected in some Seyfert II galaxies \citep{2002MNRAS.331..154R},
and gave clear evidences of the existence of BLR in type II AGN. But for the
majority of type II AGNs, the extinction of BLRs at near-IR band are still too
high to detect BELs.  A promising solution is to measure the line width of the BELs
at longer wavelengths. Proper motions and radial velocities of H$_2$O
mega-maser spots can be used to deduce a high precision central mass, such as
for the type II galaxy NGC 4258  \citep{1995Natur.373..127M,1999Natur.400..539H}.
But because of the limitation of sensitivity, this method can not be used for
a large sample of type II AGN.

Being almost free of extinction, the   radio recombination lines (RRLs) of
hydrogen, which have been detected in the Galactic H{\sc ii} regions 
\citep{1965Sci...150..339H}  and external galaxies  \citep{1977A&A....55..435S} for decades, could be an ideal choice for such studies. The $\alpha$
lines of RRLs (from energy level $N+1$ to $N$) are much stronger than the lines
with $\Delta N>1$ (e.g., $\beta,\gamma,\delta$ lines), where $N$ is the
principal quantum number. So the first choice should be the $\alpha$ lines.
The frequencies of RRLs have a wide range from centimeter to (sub-)millimeter
wavelengths, which give us many choices.

 The  Circinus galaxy, which is an extremely nearby Seyfert II galaxy at a distance
of 4 Mpc (1$''\sim 20$pc) \citep{1977A&A....55..445F}, has been detected in narrow H$\alpha$ emission   \citep{1994A&A...288..457O}  and broad  H$\alpha$  with full width half maxim (FWHM) line width of $\sim3300$ km s$^{-1}$  with polarized emission \citep{1998A&A...329L..21O}.  Near-IR Br$\gamma$ observations suggested that the narrow lines are from star formation activity surrounding the Seyfert nucleus rather than the narrow line region, while the broad line is not  detected due to obscuration even with  Br$\gamma$ \citep{2006A&A...454..481M}.  Thus, observations of millimeter RRLs toward the  Circinus galaxy with the ALMA offers a good opportunity to directly detect  emission from BEL regions of type II AGNs.

In this paper, we will describe the  observations and data reduction  in \S2,  present the main results   in \S3 and discussions in \S4, make the brief summary  in \S5.

\section{Observations and data reduction}

The observations were taken with the ALMA 12-m array on March 04 and April 29,
2017 (project number: 2015.1.00455.S, PI: Junzhi Wang).  Band 4 receivers are
used to cover H35$\alpha$ ($f_{\rm rest}$= 147.047 GHz) and H36$\alpha$ ($f_{\rm
rest}$= 135.286 GHz), in the upper side band (USB) and lower side band (LSB),
respectively.  Each sideband was covered with two spectral line windows (SPW)
of 1.875 GHz band width each. The two SPWs have an overlap of 0.18 GHz, to
ensure a maximum bandwidth coverage and consistent response in the overlap
frequency range between two SPWs. It provides about 3.5 GHz  useable frequency coverage after removing edge channels, for both LSB and USB.  The 3.5 GHz frequency coverage corresponds to $\sim$$7140$ km s$^{-1}$ for H35$\alpha$ and  $\sim$$7760$ km s$^{-1}$ for H36$\alpha$, respectively.
 Observations were taken with three
executions, with 60--70 minutes observing time each.  The total on source time
is about 108 minutes with 42 antennas, which gave the longest baseline of
$\sim$ 640 m and shortest baseline of $\sim$15 m.

Ganymede and J1427-4206, J1617-5848, and J1424-6807 were adopted as the flux,
bandpass, and gain calibrators, respectively. Atmospheric calibrators were
performed with Ganymede, J1617-5848, and Circinus, before each scans on these
calibrators and targets. We calibrate the data with standard calibration
procedures with Common Astronomy Software Applications (CASA), version 4.7
\citep{2007ASPC..376..127M}.  The default outcome from the pipeline show slight flux
offset between the overlap frequency ranges in adjacent SPWs. 
Observations with Ganymede as the flux calibrator do not show such an issue. 
Therefore, we average the ALMA-monitored fluxes of the flux calibrator \footnote{\url{https://almascience.eso.org/sc/}},
J1427-4206, observed during 25 Mar. and 23 May 2016, to be 3.02$\pm$0.07 Jy,
1.89$\pm$0.09 Jy, and 1.49$\pm$0.03 Jy,  at 91.5 GHz, 233 GHz, and 343.5 GHz,
respectively.  Then we fit its spectral index to be $-0.536 \pm 0.018$ and
implement it in {\sc setjy} of pipeline. After this correction, the overlap
frequency ranges between SPWs are consistent. 

Then we combine all data with {\sc concat}, then image and deconvolve the data
with {\sc tclean} of CASA, version 6.1.2, using a natural weighting and a pixel
size of 0.2$''$. It gave a restoring beam of $1.34''\times 1.21''$ with a
position angle (PA) of $-22.0^\circ$ at the USB and $1.47'' \times 1.31''$ with a
PA of $-25.2^\circ$ at the LSB.  The rms level of the data cube is about 0.22 mJy
beam$^{-1}$ at a frequency resolution of 19.53 MHz, which is similar to the expected rms level based on ALMA sensitivity calculator.

\section{Results}

Spatially unresolved   narrow hydrogen radio recombination lines (RRLs)  H35$\alpha$ and  H36$\alpha$ are detected toward the center of Circinus galaxy.  The spectra presented  in Figure 1 and 2 are collected from the central region of $\sim3''$ in diameter.  The spectra presented here have a frequency resolution of 19.53MHz, which corresponds to 39.82 km s$^{-1}$ and 40.36  km s$^{-1}$  for H35$\alpha$ and  H36$\alpha$, respectively. The rms of the spectra  is about 0.46 mJy at the  resolution of 19.53MHz.

The central velocity of the  H36$\alpha$   narrow component is 465$\pm$10km s$^{-1}$, with the line width  (FWHM) of 236$\pm$28 km s$^{-1}$, a velocity integrated flux of 1310$\pm$123 mJy  km s$^{-1}$, and a peak flux density of about 5.2 mJy,  based on Gaussian fitting. On the other hand,  the narrow component of  H35$\alpha$ is strongly contaminated by CS 3-2, which is about 10 times stronger than that of  H35$\alpha$ (see Figure 1), assuming similar flux of  H36$\alpha$.  The continuum emissions in both LSB and USB are close to 44.5 mJy (see Figure 1). 
Several other molecular lines are also detected,  including HC$_3$N 15-14 \&16-15, CH$_3$CCH 8-7, CH$_3$CN 8-7,  as well as H$_2$CO 2$_{0,2}$-1$_{0,1}$ at the rest frequency of 145.602856 GHz, which is blended with HC$_3$N  16-15 (see Figure 1 and 2).
The  broad RRL components of H35$\alpha$ and  H36$\alpha$ are not detected (see Figure 1).  Based  on the noise level and assuming the line width is similar to that of  broad  H$\alpha$  of $\sim3300$ km s$^{-1}$ \citep{1998A&A...329L..21O}, the 3 $\sigma$ upper limit of broad RRL emissions of H35$\alpha$ and  H36$\alpha$ is 3$\times0.46\times\sqrt{40\times3300}=501$mJy km  s$^{-1}$.  The non-detection of broad RRL emissions of H35$\alpha$ and  H36$\alpha$ makes it impossible to derive the real line width of broad line region in Circinus galaxy with these two lines.

\begin{figure}
\includegraphics[angle=0,scale=.25]{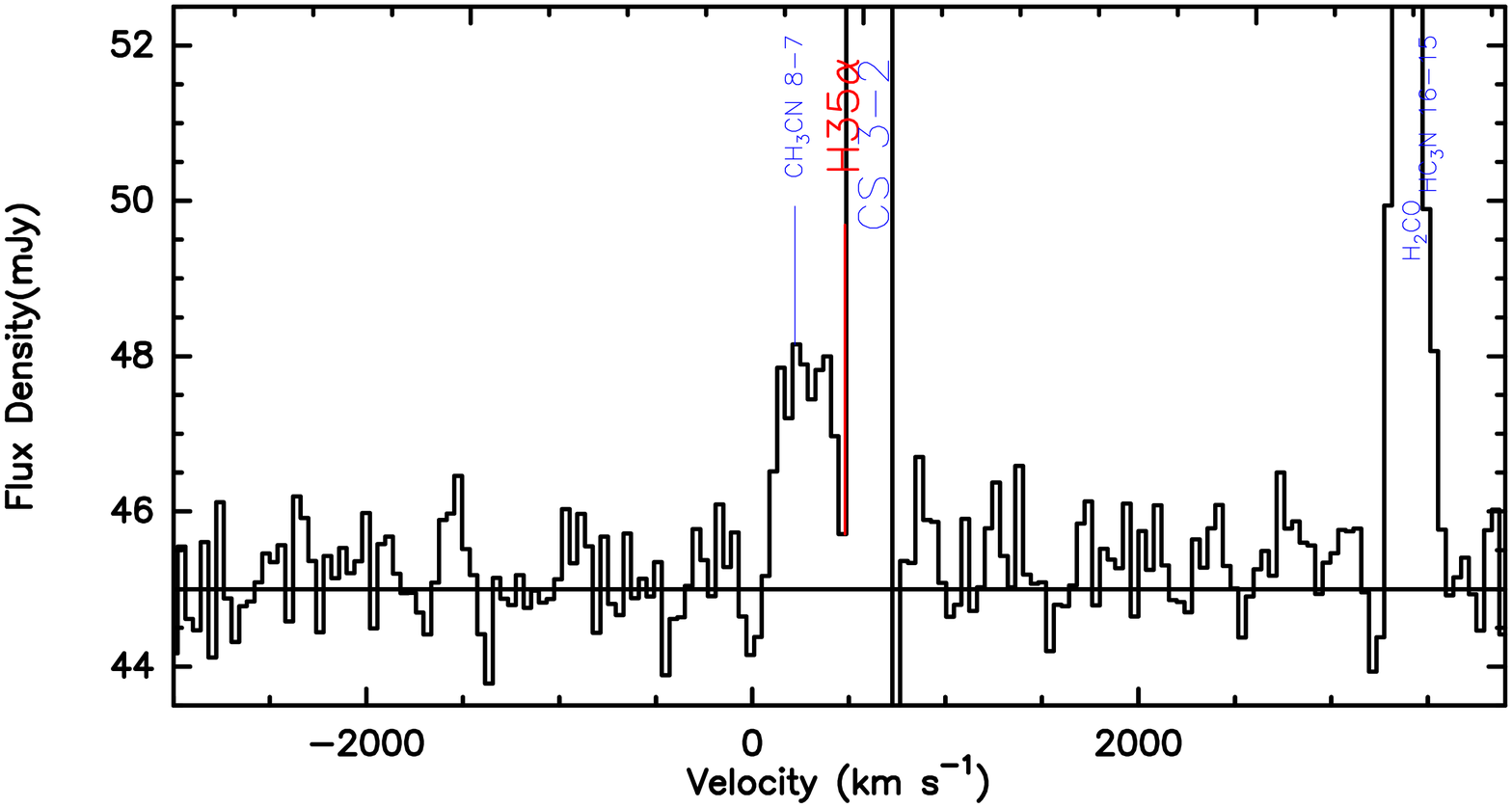}
\includegraphics[angle=0,scale=.25]{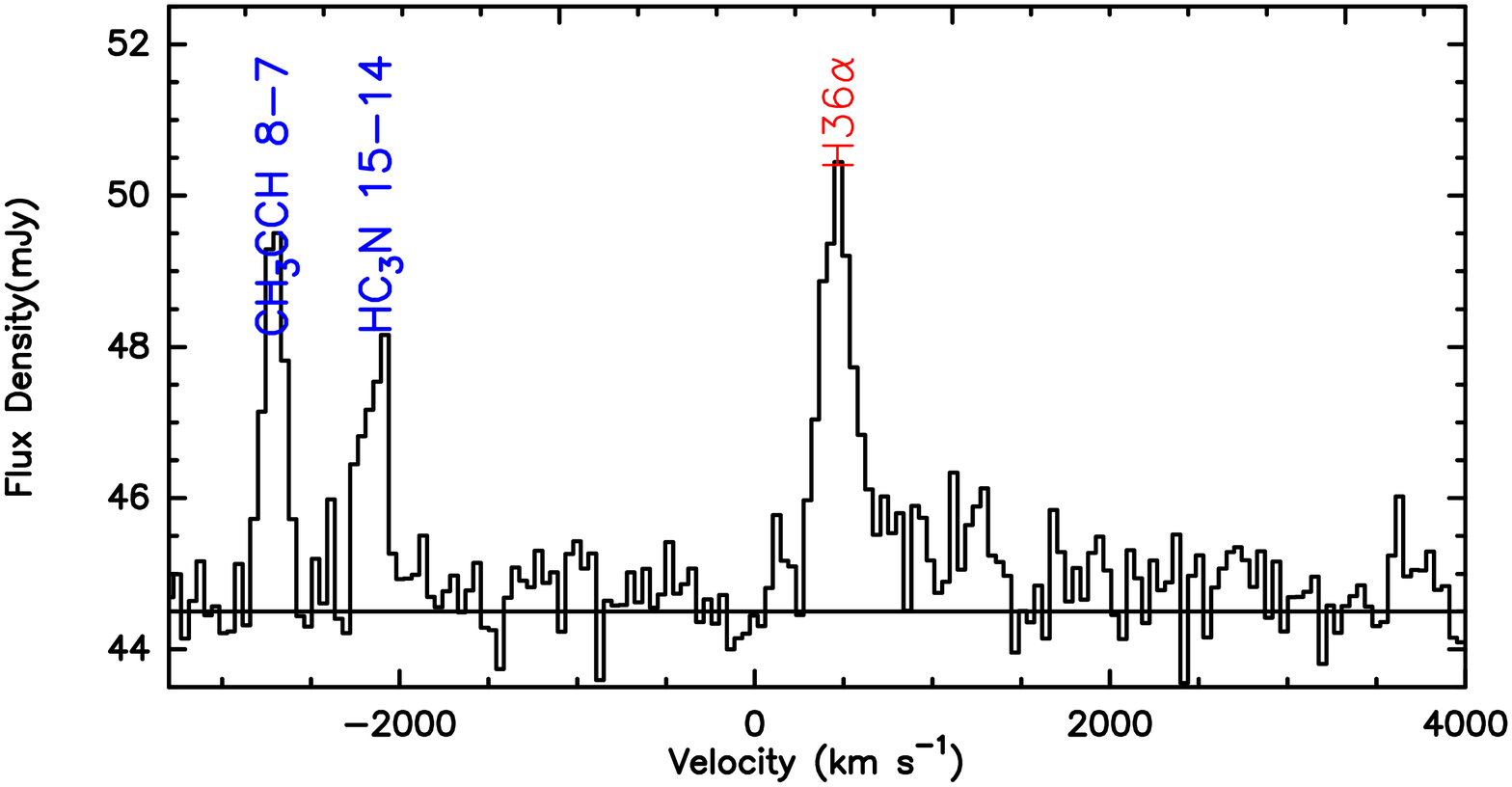}

    \caption{The spectra around the rest frequencies  of H35$\alpha$ in the top panel and  H36$\alpha$ in bottom panel, collected from the central region of $\sim3''$ in diameter. The detected lines are marked.}
    \label{fig:figure1}
\end{figure}

\begin{figure}
\includegraphics[angle=0,scale=.25]{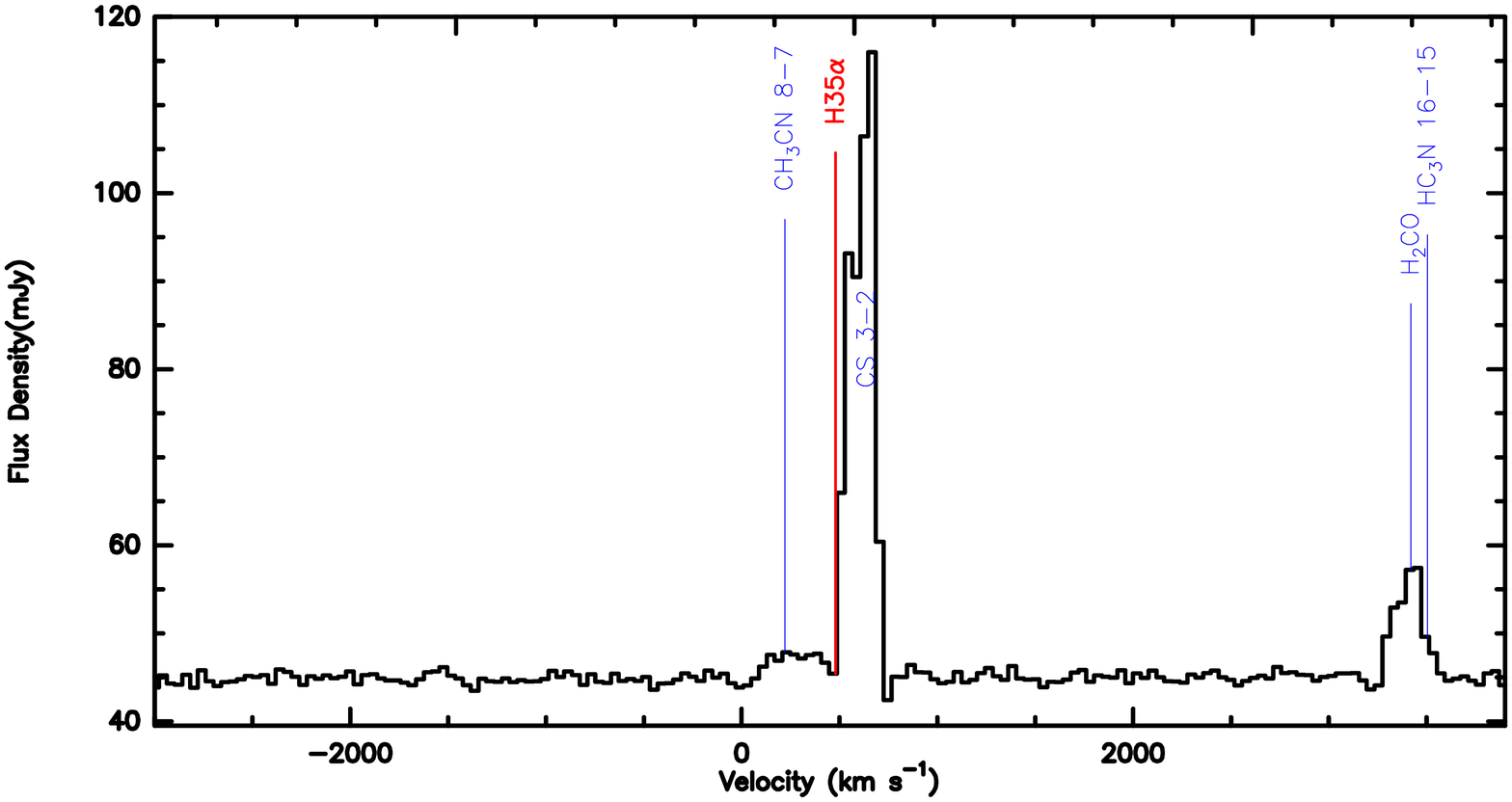}

    \caption{Full view of spectrum  for H35$\alpha$ shown in the top panel of Fig. 1.  }
    \label{fig:figure1}
\end{figure}

\section{Discussion}

\subsection{The nature of the narrow component of RRL}

Narrow RRLs in Circinus galaxy, H91$\alpha$ and H92$\alpha$, had been detected using the Australia Telescope
Compact Array (ATCA), with line width (FWHM)  of H91$\alpha$+H92$\alpha$  about  260 km s$^{-1}$ centered at 449 km s$^{-1}$, which could be explained by a collection of 50 to 10 000 H II regions with
temperatures of 5000 K  \citep{2008A&A...483...79R}.   H91$\alpha$+H92$\alpha$ emissions were mainly from the  central region, except for some contribution from   the northeast of center region with about 10$''$, while the beam size was 11.0$''\times9.4''$ with PA of -76$^{\circ}$ \citep{2008A&A...483...79R}.   H36$\alpha$ from the central 3$''$ region shown in Figure 1 has a  line width of  236$\pm$28 km s$^{-1}$ and central velocity  of 465$\pm$5km s$^{-1}$, which generally agree well with the parameters from  H91$\alpha$ and H92$\alpha$.  

The properties of ionized gas  have also been discussed  in relation to near-IR  Br$\gamma$ emission, to distinguish  the contribution from star formation near the center of Circinus galaxy or the narrow line region (NLR) of AGN. With Br$\gamma$ and H$_2$ 1-0 $S(1)$ observations with the  SINFONI on VLT UT4 at the spatial resolution of $\sim$0.22$''$ ($\sim$4.2pc)  suggested that  Br$\gamma$ emission  within the inner 1$''$ in diameter  is associated with star formation rather than the narrow line region of AGN \citep{2006A&A...454..481M}. 

Because the NLR in Circinus galaxy was estimated to be about 100 pc \citep{2018ApJ...867...49W}, which corresponds to 5$''$, and no significant  narrow H36$\alpha$ emission is found from inner 3$''$ to 5$''$ region from our data, we would like to 
 suggest that  even though the contribution from AGN NLR can not be fully excluded, 
 narrow H36$\alpha$ emission   from the inner   3$''$ ($\sim$60pc) region detected with our ALMA observation  should be mainly from star formation activity.  

%The line ratio of H91$\alpha$+H92$\alpha$  to H36$\alpha$, which is close to unity in the unit of Jy km s$^{-1}$,  indicates that some of the H91$\alpha$+H92$\alpha$ emission may be from low electron density gas with amplified emission at centimeter \citep{1996ApJ...470.1118S}. Otherwise, the line ratio under local thermal  equilibrium (LTE) condition  is $\propto\nu$   \citep{2016era..book.....C}, which means H36$\alpha$ will be more than 10 times higher than that  of 91$\alpha$ and H92$\alpha$. Since the beamsize (11.0$''\times9.4''$) of  H91$\alpha$ and H92$\alpha$ \citep{2008A&A...483...79R} is much larger than the beam of this observation and the central 2$''$ region  for collecting spectra,  the  contribution  of  H91$\alpha$ and H92$\alpha$ emission  from outside of the central 2$''$ region  may also cause the high  ratio of H91$\alpha$+H92$\alpha$ to H36$\alpha$. 

 \subsection{The absence of broad RRL emission}
 
 Broad emissions of both H36$\alpha$ and H35$\alpha$ are not detected  (see Figure 1).  There are two possible reasons for this. One is that the line emission is too weak to be detected, while the other is that the line is too broad to be covered by the bandwidth of this observation.  
Because of the extremely high electron
density ($n_e>10^9$cm$^{-3}$) of BLRs \citep{2014SSRv..183..253P}, the pressure broadening can
not be neglected, especially for the low frequency RRLs.  The ratio between 
pressure broadening and thermal Doppler effect is
1.2$\times(n_e/10^5)\times(N/92)^7$ \citep{2008ApJ...672..423K}, where $n_e$ is electron density in cm$^{-3}$ and $N$ is the principle quantum number. So, using the thermal line width (FWHM) of
$\sim21.6$ km s$^{-1}$ for the gas with an electron temperature of $T_{\rm
e}\sim$10000K, the line widths of pressure broadening are 760 km s$^{-1}$ for $N=40$ at about 99 GHz  and 360 km
s$^{-1}$ for $N=36$ at about 135 GHz, respectively, if 
$n_e$ is $10^9$cm$^{-3}$.  Since the FWHM of broad  H$\alpha$ is  $\sim3300$ km s$^{-1}$  \citep{1998A&A...329L..21O}, the pressure broadening of H36$\alpha$ and H35$\alpha$ can be neglected, even if $n_e \sim10^9$cm$^{-3}$. Thus, the non-detection of broad H36$\alpha$ and H35$\alpha$ should not be caused by the limited bandwidth that is not wide enough to cover the line.
 
 The typical size of BLRs is from several light days to  hundreds  of light days for over 30
reverberation-mapped AGNs \citep{2004ApJ...613..682P,2014SSRv..183..253P}.  However, Circinus galaxy was not included in that sample. 
 Based on continuum emission for a wavelength range from 8.0 to 13.0 $\mu$m with the MIDI interferometer at the VLTI, a  dense and warm (T $>$330 K) disk component with a radius of 0.2 pc ($\simeq$240 light days; 1pc$\simeq$3.2616 light years)
 was found \citep{2007A&A...474..837T}. BLR was thought to be surrounded by  such warm  dusty disk component. Thus, the size of BLR should be smaller than that of warm  dusty disk.  For an ionized cloud with a radius of 0.1 pc  and $n_e$ of $10^8$cm$^{-3}$, the opacity of  free-free at 135 GHz is about 1.9$\times10^4$, based on the relation of $\tau_{\nu,C}=\kappa_{\nu,C}L=9.770\times10^{-3}\frac{n_e^2}{\nu^2T_e^{1.5}}[17.72+\textrm{ln}\frac{T_e^{1.5}}{\nu}]L$ \citep{1961ApJ...134.1010O}, where $n_e$ is electron density in unit of cm$^{-3}$, $\nu$ is frequency in Hz, $T_e$ is electron temperature in $K$, and $L$ is the length of line of sight in cm.  Since the even if the size of individual clump of BLR is  $\sim$10$^{-5}$pc, free-free emission is  still optically thick at 135 GHz if $n_e$ is greater than 10$^8$ cm$^{-3}$. If  $n_e$ is 10$^{9}$cm$^{-3}$, the free-free optical depths   are 100 times  higher than that for $n_e$ of  10$^{8}$cm$^{-3}$.

 If  atomic hydrogens  in BLR that   emit radio recombination lines are under local thermal equilibrium (LTE) conditions, there should not be line emission when free-free emission is optically thick at the same frequency.   However, the level populations of atomic hydrogens can be inverted in such ionized gas clouds \citep{1996ApJ...470.1118S} and \citep{2022A&A...665A..94Z}.  Based on the model in \cite{2022A&A...665A..94Z},  the line to continuum ratio at the line peak is about 1\% for H36$\alpha$ with FWHM of  turbulence to be  3300 km s$^{-1}$, where the continuum emission is due to free-free from electrons in BLR.  Since the total continuum at 321 GHz at the nuclear  region is only  about 50 mJy \citep{2021ApJ...923..251H}, which include thermal dust emission and optically thin free-free emission, the  optically thick free-free emission from BLR should be less than 50 mJy at this frequency.  Since optically thick free-free  continuum  is $\sim\nu^2$, it should be less than 9 mJy at 135 GHz, while total continuum is about 44 mJy (see Figure 1).  It is impossible to detect  a signal weaker than 0.09 mJy, which is $\sim$1\% of the optically thick free-free emission, with our observation at the rms level of 0.46 mJy at the frequency resolution of 19.53MHz. 

The flux of optically thick free-free emission at given frequency is related to electron temperature ($T_e$) and solid angel  ($\Omega$) as $\propto T_e\times\Omega$.  $T_e$ can only vary within a very small range around 10000K and $\Omega$ is $\propto (d/D)^2$, where $d$ is the source size and $D$ is the distance to the earth. For  Circinus galaxy at the distance of 4 Mpc (1$''\sim 20$pc) \citep{1977A&A....55..445F},  if the BLR is 0.2 pc in diameter (i.e., 0.1pc in radius), the view angle is 0.01$''$, which gives the conversion factor from K to Jy is $\simeq1.7\times10^{-6}$. Thus, the free-free continuum from  BLR will be 17mJy at 135 GHz if  the size is 0.1pc in radius.  Since the estimated upper limit of optically thick free-free emission at 135GHz is  9 mJy, the radius of BLR in  Circinus should be less than 0.07 pc.

  \subsection{Future prospects}

With the non-detection of broad H35$\alpha$ and H36$\alpha$  in  Circinus galaxy at the distance of 4 Mpc \citep{1977A&A....55..445F},  as one of the closest type II AGNs, it is hopeless to detect such emission at similar frequency with ALMA.  Since the line to continuum ratio is about 1\% for H36$\alpha$ (see \S4.2), if we expect to detect the signal with line peak of $\sim0.5$ mJy,  the optically thick free-free emission at 135 GHz should be $\sim$50 mJy, which requires the  radius to be $\sim$0.17pc at the distance of 4 Mpc,  or $\sim$0.54pc for 40 Mpc, which is larger than the maxim size of BLR of reverberation-mapped AGNs \citep{2004ApJ...613..682P,2014SSRv..183..253P}, even if the filling factors of BLRs are unity. 

Sub-millimeter RRLs, which are stronger than that at millimeter band,  can be another choice for sensitivity consideration. We use H26$\alpha$ at 353.6 GHz as an example to do such estimation. For an ionized cloud with radius of 0.1 pc, the optical depth of free-free emission at this frequency is about 2300, if $T_e=10000$K. The line to continuum ratio at line peak is $\sim$12\% if line width is 3300 km s$^{-1}$.  If the expected line  peak of  H26$\alpha$ is 0.5 mJy, the required  optically thick free-free continuum will be only  4.2 mJy at 353.6 GHz  or 0.61 mJy at 135GHz, which corresponds to  the radius of $\sim$0.019pc at the distance of 4 Mpc, or 0.059 pc for 40 Mpc. The  capability of detection will be  better with the higher frequency of RRL. However, the frequency  coverage of ALMA with maxim capability of $\sim$3.7GHz can not provide enough velocity coverage to detect broad RRL emission. The corresponding velocity coverage at 353.6 GHz for H26$\alpha$ is only $\sim$3140 km s$^{-1}$.  Non-detection of such broad emission of H26$\alpha$ in the famous seyfert  II galaxy  NGC 1068  had been reported in the literature \citep{2016MNRAS.459.3629I}, which was mainly due to the limited velocity coverage at sub-millimeter wavelength.  

Thus, even though it can not be done with current (sub-)millimeter facilities, observing RRLs at sub-millimeter band with ALMA after the upgrade of wide bandwidth, or at millimeter band with next generation Very Large Array (ngVLA), are possible to directly detect BLRs from local AGNs in the future.

\section{summary}% and conclusion remarks }

With ALMA  observations of H35$\alpha$ and  H36$\alpha$   toward  Circinus galaxy, we obtained  upper limit of these lines  from broad line regions, which are thought to be with line widths (FWHM) of thousands of km s$^{-1}$,   in type II AGNs. Narrow  H36$\alpha$ emission with line width (FWHM) of   236$\pm$28 km s$^{-1}$  at $\sim$ 10 $\sigma$ level, which is  thought to be mainly   from star forming regions around the nuclear region, is detected.    Narrow H35$\alpha$, contaminated by CS 3-2, is also detected.

With the estimation by free-free continuum emission and line to continuum ratio at this frequency, we conclude that broad H35$\alpha$ and H36$\alpha$ in  Circinus galaxy can not be detected with ALMA due to limited sensitivity. Since Circinus galaxy is one of the nearest type II AGN, it is hopeless to detect broad H35$\alpha$ and H36$\alpha$  for other local AGNs. 
 On the other hand, sub-millimeter RRLs, which can be dozens stronger than  H36$\alpha$, can be used to direct detect BLR in local AGNs with the ALMA,  after its  backend is  upgraded to be several times better than current capability.

\section*{Acknowledgements}

This work is supported by  the Natural Science Foundation of China under grants of  12173067. This paper makes use of the following ALMA data: ADS$/$JAO.ALMA${\#}$2015.1.00455.S. ALMA is a partnership of ESO (representing its member states), NSF (USA) and NINS (Japan), together with NRC (Canada), MOST and ASIAA (Taiwan), and KASI (Republic of Korea), in cooperation with the Republic of Chile. The Joint ALMA Observatory is operated by ESO, AUI/NRAO and NAOJ.

%We thank the anonymous referee for helpful suggestions to improve the manuscript.  
%This work is supported by  the National Natural Science Foundation of China grant 11590783, and U1731237.  This study is based on observations carried out under project number 066-19, 186-18 and 058-17  with the IRAM 30-m telescope. IRAM is supported by INSU/CNRS (France), MPG (Germany) and IGN (Spain).  This research has made use of the NASA/IPAC Extragalactic Database, which is funded by the National Aeronautics and Space Administration and operated by the California Institute of Technology.

\section*{Data availability}
The original   data observed with ALMA  can be accessed by ALMA archive system at  http://almascience.eso.org/asax/. If anyone is interested in the  calibrated data, please contact  Junzhi Wang at  junzhiwang@gxu.edu.cn.

%%%%%%%%%%%%%%%%%%%%%%%%%%%%%%%%%%%%%%%%%%%%%%%%%%

%%%%%%%%%%%%%%%%%%%% REFERENCES %%%%%%%%%%%%%%%%%%

% The best way to enter references is to use BibTeX:

%\bibliographystyle{mnras}
%\bibliography{example} % if your bibtex file is called example.bib

% Alternatively you could enter them by hand, like this:
% This method is tedious and prone to error if you have lots of references

%%%%%%%%%%%%%%%%%%%%%%%%%%%%%%%%%%%%%%%%%%%%%%%%%%

%%%%%%%%%%%%%%%%% APPENDICES %%%%%%%%%%%%%%%%%%%%%

%\appendix

% Don't change these lines
\bsp	% typesetting comment

\label{lastpage}
\end{document}